\documentclass[conference]{IEEEtran}
\IEEEoverridecommandlockouts

\usepackage{cite}
\usepackage{amsmath,amssymb,amsfonts}
\usepackage{algorithmic}
\usepackage{graphicx}
\usepackage{textcomp}
\usepackage{xcolor}
\usepackage{enumitem}
\usepackage[hidelinks]{hyperref}
\usepackage{url}
\def\BibTeX{{\rm B\kern-.05em{\sc i\kern-.025em b}\kern-.08em
    T\kern-.1667em\lower.7ex\hbox{E}\kern-.125emX}}

\begin{document}

\newcommand{\etal}{~et~al.}
\newcommand{\IIPEcosphere}[0]{IIP-Ecosphere}
\newcommand{\OPCUA}[0]{OPC UA}
\newcommand{\oktoflow}[0]{oktoflow}
\newcommand{\idtaSpecs}[0]{IDTA-specs}
\newcommand{\idtaSpec}[0]{IDTA-spec}
\newcommand{\semanticId}[0]{semanticId}
\newcommand{\semanticIds}[0]{semanticIds}
\newcommand{\idShort}[0]{idShort}
\newcommand{\idShorts}[0]{idShorts}

% For commenting:
\definecolor{black}{rgb}		{0.0, 0.0, 0.0}
\definecolor{white}{rgb}		{1.0, 1.0, 1.0}
\definecolor{yellow}{rgb}		{1.0, 1.0, 0.8}
\definecolor{red}{rgb}			{0.6, 0.0, 0.2}
\definecolor{blue}{rgb}		{0.0, 0.2, 0.5}
\definecolor{green}{rgb}		{0.6, 0.8, 0.8}
\definecolor{dark_green}{RGB} {0, 140, 0}
\definecolor{gold}{rgb}		{0.6, 0.4, 0.1}
\definecolor{grey}{RGB}{0,0,0}
\definecolor{Gray}{gray}{0.8}
\definecolor{MediumGray}{gray}{0.9}
\definecolor{LightGray}{gray}{0.98}
\definecolor{LightCyan}{rgb}{0.88,1,1}
\definecolor{purple}{RGB}{128,0,128}
\definecolor{slblue}{RGB}{47, 60, 105}
\definecolor{orange}{RGB}{255,165,0}
\definecolor{Gray}{gray}{0.85}
\definecolor{light_green}{RGB}{226, 240, 217}

\newcommand{\numP}[1]{\textcolor{purple}{[#1 p]}}

\newcommand{\comHE}[1]{\textcolor{purple}{\textbf{\large [}\colorbox{yellow}{\textbf{Holger:}}{\small #1}\textbf{\large ]}}}
\newcommand{\comAW}[1]{\textcolor{purple}{\textbf{\large [}\colorbox{yellow}{\textbf{Claudia:}}{\small #1}\textbf{\large ]}}}

\newcommand{\nonbreakingHyphen}[0]{\mbox{-}}

\title{Model-driven realization of IDTA submodel specifications: The good, the bad, the incompatible?\\}

\author{\IEEEauthorblockN{Holger Eichelberger}
\IEEEauthorblockA{\textit{Institute for Computer Science} \\
\textit{University of Hildesheim}\\
Hildesheim, Germany \\
eichelberger@sse.uni-hildesheim.de}
\and
\IEEEauthorblockN{Alexander Weber}
\IEEEauthorblockA{\textit{Institute for Computer Science} \\
\textit{University of Hildesheim}\\
Hildesheim, Germany \\
weber@sse.uni-hildesheim.de}
}

\maketitle

\begin{abstract}
 Asset Administration Shells are trending in Industry 4.0. In February 2024, the Industrial Digital Twin Association announced 84 and released 18 AAS submodel specifications. As an enabler on programming level, dedicated APIs are needed, for which, at this level of scale, automated creation is desirable. In this paper, we present a model-driven approach, which transforms extracted information from IDTA specifications into an intermediary meta-model and, from there, generates API code and tests. We show we can process all current IDTA specifications successfully leading in total to more than 50000 lines of code. However, syntactical variations and issues in the specifications impose obstacles that require human intervention or AI support. We also  discuss experiences that we made and lessons learned. 
\end{abstract}


\section{Introduction}\label{sect-introduction}

Standards are a corner stone for interoperability in Industry 4.0/IIoT. The Asset Administration Shell (AAS) is a new standard (IEC 63278-1 ED1) aiming to increase the interoperability of industrial assets and products. The Industrial Digital Twin Association (IDTA) performs valuable work on the AAS standard and on submodel formats for different purposes, e.g., nameplates or bill-of-material structures. In February 2024, 18 IDTA submodel specifications were published\footnote{\url{https://industrialdigitaltwin.org/content-hub/teilmodelle}} and in total 84 were announced. While individual applications may rely on selected specifications, an AAS tool or an Industry 4.0 platform may have to offer support for all submodel specifications. Manually realizing individual specifications, e.g., as Application Programming Interfaces (APIs), may be, depending on size and complexity, a tedious yet feasible approach. However, for an Industry 4.0 platform, manually providing a consistent realization of such a large and evolving set of specifications does not scale. One resort could be automation, e.g., through model-driven engineering~\cite{BCW12}, which can automatically map the specific concepts of the specifications onto the generic implementation of AAS in a framework like Eclipse BaSyx\footnote{\url{https://eclipse.dev/basyx/}}.

In this context, we ask the following research questions: 1) Can we create a model-driven engineering approach for IDTA submodel specifications (\idtaSpecs{})? 2) Can the actual \idtaSpecs{} be used as a basis for an automated realization? 3) How many of the currently available \idtaSpecs{} can be treated to what degree in an automated manner? 4) Can we identify issues or improvements in the automated process?

As contributions, we present a model-driven approach which employs code generation to derive a functional submodel API per \idtaSpec{}. We evaluate the individual steps of our approach on the two currently predominant \idtaSpec{} formats, namely PDF and AASX, for 18 \idtaSpecs{} and one submodel specification draft. To improve the results and to increase tolerance, we consider Artificial Intelligence (AI). We conclude that the PDF specification files contain more consistent and relevant information than the machine-readable AASX files and that we are able to automatically derive API implementations and related testing code (in total more than 50 KLOC) for all employed specifications. Based on these results, we discuss lessons learned and suggestions.

We realized our approach in the context of the Open Source Industry 4.0 platform \oktoflow{}\footnote{\url{https://oktoflow.de/}} (former \IIPEcosphere{} platform~\cite{EichelbergerNiederee23}), which allows to derive even more code, e.g., for data transport, serialization or data connectors and, thus, provides an environment for future work. Although we validate our work in this context, the approach can similarly be realized with other model-driven infrastructures, e.g., based on the Meta Object Facility (MOF) or Eclipse Ecore. 

Structure of the paper: In Section~\ref{sect-approach}, we introduce our model-driven approach and it's technical realization in Section~\ref{sect-realization}. Our main focus is on the evaluation in Section~\ref{sect-evaluation}, which is the basis for the lessons-learned discussion in Section~\ref{sect-lessons}. In Section~\ref{sect-related}, we elaborate related work and in Section~\ref{sect-conclusion} we conclude and indicate future work.

\section{Approach}\label{sect-approach}

In this section, we briefly introduce the concept of AAS as well as the goals and the main steps of our approach. 

An \textbf{AAS} is a hierarchical model, structuring an Asset and its "shell" in terms of so called submodels, which, in turn, consist of submodel elements. Among others, a submodel element can be a typed property, a data element (e.g., file, BLOB, range), an operation, a collection of such elements as well as a reference to or a relationship among submodel elements. Besides a type and a value, AAS elements can be detailed by a multi-language description or a semantic identifier (\semanticId{}), which can point to a local concept description or a global catalogue such as ECLASS\footnote{\url{https://eclass.eu/}} or IEC Common Data Dictionary CCD\footnote{\url{http://cdd.iec.ch/}}. In more recent AAS versions, also lists of submodels or submodel elements are available. An AAS can be served online or provided in terms of JSON or the more encompassing AASX format, which packages an XML serialization of the AAS with accompanying resources and documents into a standardized folder structure.

As we aim at automatically deriving program code for \idtaSpecs{}, we consider the following \textbf{goals}:
\begin{enumerate}[label=G\arabic*]
  \item\label{req-gen} Turn the structure of an \idtaSpec{} automatically and in uniform manner into API source code easing the programmatic creation of compliant AAS, their submodels and (typed) submodel elements.
  \item\label{req-constraints} Validate the created structures based on constraints in the specifications, in particular cardinalities for properties.
  \item\label{req-addInfo} Consider information that indicates, e.g., how names of properties with multiple values are constructed, whether the name of a submodel (element) may be changed by the user, alternative \semanticIds{} can be used or values belong to an (extensible) enumeration. 
  \item\label{req-resolve} Resolve types to increase consistency and consider "imported" specifications via their \semanticId{} to facilitate reuse. Also consider "reuse" mechanisms, which, e.g., "copy" a part of a specification into multiple other parts.
  \item\label{req-test} Validate the generated API code through unit tests based on examples given in the \idtaSpecs{}.
\end{enumerate}

\begin{figure*}[tbp]

\centerline{\includegraphics[page=3,trim={0.0cm 9cm 0.2cm 0.1cm},clip, scale=0.55]{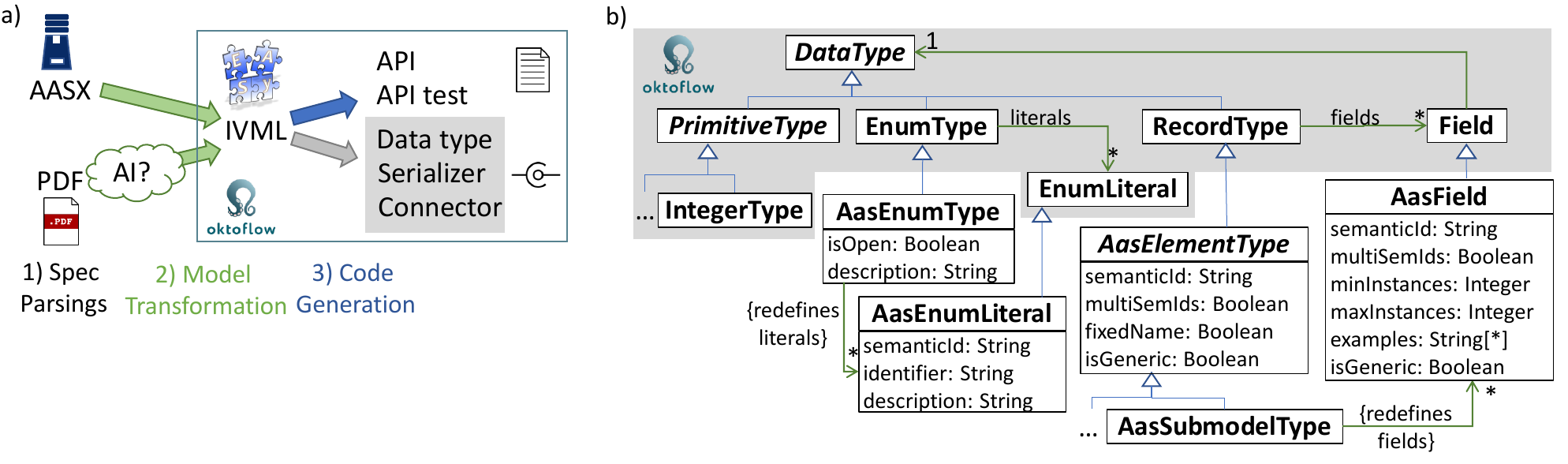}}

\caption{Approach: a) main steps - parsing, model transformation and generation, b) simplified intermediary meta model aligned with \oktoflow{}.}
\label{fig_approach}
\label{fig_meta}
\end{figure*}

As illustrated in Figure~\ref{fig_approach} a), our \textbf{approach} consists of three steps, namely 1) parsing \idtaSpecs{} from PDF or AASX, 2) transforming the extracted information to an intermediary model that is better suited for code generation, and 3) processing  the intermediary model to obtain API code for the specified AAS structure and elements.

In the first step, we parse the IDTA input. One primary source are the serialized AAS models provided by the IDTA for all \idtaSpecs{} in \textbf{AASX} file format. AASX may contain a template structure of a specification, e.g., defining names, types, cardinalities and nesting of submodel elements. An alternative could be the AAS JSON format, which is provided by the IDTA only for one specification. A second, less obvious source is the \textbf{PDF} specification file. In particular for early adopters of submodel drafts, where no AASX file may yet be provided, the PDF may be the only source. Besides explanatory text, the hierarchical structure of an AAS is defined here in terms of tables. For submodels and element collections, these tables consist of two columns (header, value). For submodel elements, the tables have four columns. In these tables, the first two rows state the headers (idShort, \semanticId{}/description, value type/example and cardinality) while each following row defines a contained submodel element according to the headers. Although not intended to be machine-readable, existing libraries or services may extract the tables and the contained information. 

In the second step, the extracted information is transformed into an intermediary model. In model-driven engineering~\cite{BCW12}, a given input model for a model-transformation often may not be the best starting point for code generation. Frequently, input models are transformed into an \textbf{intermediary meta-model} that may include additional information (\ref{req-addInfo}) or resolved model imports (\ref{req-resolve}). Such a model augmentation may happen automatically or involve other sources like humans, e.g., to add specific examples for tests or to group related properties  into one API operation, such as status and change timestamp. To apply the approach in a larger context and to enable later validation in practical settings, we align the intermediary meta model with the \oktoflow{} platform~\cite{EichelbergerNiederee23}. On the one side, we can reuse the generation infrastructure of \oktoflow{}, which already addresses other relevant standards like OPC UA or MQTT. On the other side, we can use the extracted models for \idtaSpecs{} to support the standard-compliant development of AAS-based \oktoflow{} applications.  

\oktoflow{} utilizes for modeling and code generation the Integrated Variability Modeling Language (\textbf{IVML}) implemented by the EASy-producer toolset~\cite{EichelbergerQinSizonenko+16}. Besides value propagation to increase consistenty, one helpful ability of IVML for evolving specifications is versioning of models and imports. In other words, IVML models for \idtaSpecs{} explicitly declare the respective specification version and, when referring to another specification, the importing model defines, which version use. Although intentionally not identical, IVML is rather similar to MOF and Eclipse Ecore, i.e., our approach can be transferred to realizations based on other model-driven infrastructures. 

To allow for the modeling of data and formats for various  standards including OPC UA, the oktoflow meta-model is based on a generic notion of data types, depicted as \texttt{DataType} in Figure~\ref{fig_meta} b). oktoflow data types can be primitives (\texttt{PrimitiveType}) like Integer (\texttt{IntegerType}) or String, enumerations with literals (\texttt{EnumType}), typed lists, or records (\texttt{RecordType}) consisting of typed fields (\texttt{Field}). Our intermediary meta-model for AAS extends the oktoflow types. For example, submodels, submodel element collections or entities become records with specifically refined fields. In the context of AAS, an \texttt{AasField} indicates by its type whether it is, e.g., a property or a data element. Further, types, fields and enumerations hold additional information as targeted by \ref{req-addInfo}. However, to represent the idShort names from the \idtaSpecs{},  we had to relax some of the stricter \oktoflow{} name constraints for AAS types and fields.

In the third step, we turn an intermediary model into \textbf{API code}. We focus here on uniform API code (\ref{req-gen}) that can be used to create or query standard-compliant IDTA submodels. Before the creation of such an AAS is completed, the API validates the AAS with respect to specification constraints (\ref{req-constraints}). Besides API code, we generate also unit tests (\ref{req-test}) based on the examples in the submodel specifications as well as a build description, which compiles the API code and executes the unit tests. As frequently done in model-driven engineering, we employ templates for the code generation, which are instantiated by information in the intermediary model. Additional code generation realized by \oktoflow{}, e.g., for data transport or data connectors, as indicated in the shaded area in Figure~\ref{fig_approach} a) can also be applied to our intermediary AAS models, but is out of scope here. 

\section{Realization}\label{sect-realization}

For realizing our approach, we developed the extended \oktoflow{} meta-model from Section~\ref{sect-approach}, a parser/model-transformer for each format (PDF, AASX), and code generation templates. Below, we briefly discuss the realization.

To achieve a better integration with \oktoflow{}, we realized both input model transformers in Java\footnote{Models and implementations are on \url{https://github.com/iip-ecosphere/platform/}, APIs are generated during tests.}. For the \textbf{PDF input}, we initially experimented with several libraries including Apache PDF box, E-ICEBLUE Spire.PDF, jPDF Tweak and  tabula-java. However, none of the libraries was able to completely extract the relevant tables from the \idtaSpecs{}. Similarly, experiments with Python utilising a combination of pdfminer and pdfplumber, haystacks PDFToTextConverter
or PyPDF2 failed. Among them, the most promising approach was PyPDF2 for text-only extraction, which covered the entire text, but lost the table structures making it hard for further processing. Ultimately, we resorted to the cloud service SmallPdf\footnote{\url{https://smallpdf.com/}}, which converts PDF to Excel tables. We validated the output for several \idtaSpecs{} and found that the results are rather good, but as PDF is a primarily a print format, the Excel files may contain additional line breaks. This affects the detection of information fragments, e.g., the ending of a \semanticId{} in form of a URL followed without clear separator in the same table cell by a textual description. Further, SmallPDF tends to split single table rows in certain cases into multiple ones. Both issues complicate the extraction process as, e.g., heuristics are needed to determine the end of information or when to re-join table rows. The Excel input also has advantages as it, e.g., contains information on font sizes, which allows to correctly exclude footnotes in \idtaSpecs{}. Finally, we based the parser on Apache POI\footnote{\url{https://poi.apache.org/}}, a library for Office formats. In summary, the PDF-based realization does not allow for a completely automated translation chain, but it enables manual correction of discussed translation issues, removal of review line numbers in draft specifications as well as fixing of semantic inconsistencies in \idtaSpecs{}, which are detected during model translation.

For \textbf{AASX-based input}, we aimed at using Eclipse BaSyx. However, initial experiments showed that BaSyx (in \oktoflow{} supported versions 1.0 and 1.3) is not able to read all AASX files. In particular BaSyx, does not process specifications published after November 2023, which may already be based on AAS specification v3.0. To be able to read all AASX files, we developed an own, simple, XML DOM parser based on the default Java library. However, due to the two major AAS versions used by IDTA (v2.0 and v3.0) and due to various format differences regarding cardinalities, concept descriptions, examples, or XML prefixes, the realization was not as straight forward as expected and we had to handle more variants that we initially planned for.

The subsequent \textbf{model-transformation} to the IVML-based meta-model relies on a set of shared classes representing the extracted information. After parsing, the PDF/Excel-transformer rewrites the extracted information to apply "reuse" mechanisms (\ref{req-resolve}) that are already resolved in AASX files. In the simple form, reuse is indicated by an (illegal) \idShort{} name consisting of multiple names separated by "or", i.e. multiple similar property specifications in one table row. In the more complex form used in recent specifications, type fragments can be merged akin to aspect-oriented programming into multiple target types. Further, both transformers ensure the uniqueness of names and type definitions and resolve \semanticIds{} (\ref{req-resolve}) to establish imports among specifications and to fix issues with value types that are missing in the \idtaSpecs{}. Finally, the extracted information is exported as of IVML models. 

Before designing the \textbf{code generation}, we manually implemented 5 IDTA specifications to identify a uniform style and a supporting design. The support classes ease the code generation, as they implement reusable, helpful functions, e.g., to tolerantly turn example values into test values of a given type. The development of the code generation templates was rather straight forward, except for the alternatives implied by \ref{req-addInfo}: Depending on the \idtaSpecs{}, additional API parameters determining the actual idShort of a submodel or the actual \semanticId{} of a submodel element must be generated. Further, generation schemas for open enumerations or multi-valued return values have to be realized. 

In more details, we separated the Java target code into AAS creation, AAS access and testing. We create  
\begin{enumerate}
\item an API class for AAS creation according to the nested structure of a submodel specification (\ref{req-gen}). We represent each defined submodel (element) in nested builder-style\footnote{\url{https://en.wikipedia.org/wiki/Builder_pattern}}. A builder allows for creating complex structures through chained method calls, which implicitly set fixed \semanticIds{}, descriptions etc. and, thus, encapsulates details of the specification and the underlying AAS framework. In the last chained method call, the builder validates the imposed conditions (\ref{req-constraints}) and  returns a submodel (element) instance. If present, we utilize the descriptions from the specification as API documentation (JavaDoc).
\item an API class for AAS access, which consists of getters for typed property values as well as getters for AAS elements to allow for changing their values. Depending on the cardinality, the result can be multi-valued. As for the creation API, we use the specification descriptions to generate JavaDoc code documentation.
\item a test class, which addresses both, AAS creation and access API. In a first pass, a test builds required AAS structures through the creation API using example values from the specification or, as fallback, default values. In a second pass, the test reads these structures through the access API and compares the returned values with expected values used during the creation. As some of the \idtaSpecs{} define recursive structures, the tests must contain mechanisms to prevent accidental endless recursions.
\end{enumerate}

During code generation, we also create a build specification, which compiles, tests and packages the generated classes.

\section{Evaluation}\label{sect-evaluation}

For the evaluation of our approach\footnote{Evaluation material: \url{https://doi.org/10.5281/zenodo.11615266}}, we obtained as subjects all IDTA submodel specifications published mid of February 2024 as well as one draft specification. For short, referenced by the specification numbers, the subjects target contact information and tracing (02002, 02010), nameplates (02001, 02003, 02006, 02007), documentation handover (02004), hierarchical structures (02011), interfaces (02015, 02016, 02017), reliability (02013), safety (02014), data formats and models (02001, 02005, 02008 02012, 02021) and the Product Carbon Footprint (02023, draft 2023-01-24). 

\begin{table*}[htbp]
\centering
\caption{Subjects: IDTA Specifications published in February 2024 based on the PDF files.}
\setlength{\tabcolsep}{0.5em} % for the horizontal padding
\begin{tabular}{ |l|c|c|c|c|c|c|c|c|c|c|c|c|c|c|c|c|c|c|c| } 
\hline
\textbf{Aspects}&\rotatebox[origin=c]{90}{\textbf{02001}}&\rotatebox[origin=c]{90}{\textbf{02002}}&\rotatebox[origin=c]{90}{\textbf{02003}}&\rotatebox[origin=c]{90}{\textbf{02004}}&\rotatebox[origin=c]{90}{\textbf{02005}}&\rotatebox[origin=c]{90}{\textbf{02006}}&\rotatebox[origin=c]{90}{\textbf{02007}}&\rotatebox[origin=c]{90}{\textbf{02008}}&\rotatebox[origin=c]{90}{\textbf{02010}}&\rotatebox[origin=c]{90}{\textbf{02011}}&\rotatebox[origin=c]{90}{\textbf{02012}}&\rotatebox[origin=c]{90}{\textbf{02013}}&\rotatebox[origin=c]{90}{\textbf{02014}}&\rotatebox[origin=c]{90}{\textbf{02015}}&\rotatebox[origin=c]{90}{\textbf{02016}}&\rotatebox[origin=c]{90}{\textbf{02017}}&\rotatebox[origin=c]{90}{\textbf{02021}}&\rotatebox[origin=c]{90}{\textbf{02023}}&\rotatebox[origin=c]{90}{\textbf{2023-01-24}}\\ 
\hline
\textbf{version}&1-0&1-0&1-2&1-2&1-0&2-0&1-0&1-1&1-0&1-0&1-0&1-0&1-0&1-0&1-0&1-0&1-0&0-9&-\\
\textbf{publication}&8/22&5/22&8/22&3/23&12/23&10/23&8/23&3/23&10/23&04/23&11/23&11/22&11/23&4/22&4/23&1/24&1/24&11/23&-\\
\textbf{formats}&pac&pa&pa&pa&paj&pa&pa&pa&pa&pa&pa&pa&pa&pac&pa&pa&pa&pa&p\\
\hline
\textbf{submodels}&2&1&1&1&1&1&1&1&1&1&1&1&1&1&1&1&1&1&1\\
\textbf{types}&7&6&6&5&16&13&6&12&7&3&7&3&7&6&5&23&29&5&5\\
\textbf{elements}&15&36&18&27&69&85&37&48&26&11&50&14&31&17&16&108&217&31&31\\
\textbf{operations}&0&0&0&0&0&0&0&3&0&0&0&0&0&0&0&0&0&0&0\\
\textbf{import}&-&-&-&-&i&i&i&-&i&-&-&-&-&-&-&-&-&-&-\\
\textbf{reuse}&x&-&-&-&-&-&-&-&-&-&-&-&-&-&-&x&-&-&x\\
\hline
\textbf{notes}&x&x&x&x&-&x&x&x&x&x&x&-&-&x&-&x&-&x&x\\
\textbf{cardinality}&1..*&[1..*]&1..*&1..*&1..*&[1..*]&[1..*]&1..n&1..*/n&1..*&1..*&1..*&1..*&1..*,*&1..*,*&1..*&1..n&1..n&1..*\\
\textbf{idShort}&ca&c&ca&c&-&ca&-&v&c&ct&-&-&-&c&c&ct&a&c&c\\
\textbf{descriptions}&-&-&-&-&-&-&-&pd&-&-&-&-&-&-&-&-&fn&f&-\\
\textbf{multi-lang}&\textit{l},&@\textit{l}&@\textit{l}&@\textit{l}&-&@\textit{l}&x&@\textit{l}&@\textit{l}&-&\textit{l},&-&-&x&x&-&@\textit{l}:&-&-\\
\textbf{unit}&-&-&[\textit{u}]&-&-&Unit:\textit{u}&-&[\textit{u}],\textit{u}&-&-&-&[\textit{u}]&[\textit{u}]&-&-&-&u&[\textit{u}]*&[\textit{u}]*\\
\hline
\end{tabular}
\label{tab_subjects}
\end{table*}

Table~\ref{tab_subjects} summarizes selected general as well as differentiating characteristics of the subjects. General characteristics are the document \textit{version}, \textit{publication} month, the provided \textit{formats} (\underline{P}DF, \underline{A}ASX, \underline{J}SON, with initially \underline{c}orrupted PDFs preventing data extraction) as well as the number of defined structures including \textit{submodels}, \textit{types} (collections, entities, lists) and \textit{elements} (properties, data elements, relations, references). Only specification 02008 defines \textit{operations} as "future outlook", which we do not further address here. It is also interesting that 4 \idtaSpecs{} \textit{import} structures from the contact information specification (02002) and three specifications utilize different mechanisms to reuse type specifications or fragments. 

More than 73\% of the \idtaSpecs{} include in the \idShort{}/description \textit{notes} with relevant information for \ref{req-addInfo}. All specifications define the \textit{cardinality} of their properties, however, using 4 different notations as exemplified by the used (n)one-to-many form in Table~\ref{tab_subjects}, while specification 02010 even mixes two forms. In some specifications, a multi-valued property is broken down into multiple properties, which is indicated by a generic (illegal) \textit{\idShort{}}, e.g., \texttt{prop\{00\}} for \texttt{prop01}, \texttt{prop02}, etc. Besides this \underline{c}ounting form, we also found further forms also indicated by curly braces, e.g., containing "keywords" (\underline{a}rbitrary, \underline{v}ariable) or descriptive \underline{t}ext. While most \idtaSpecs{} use plain text \textit{descriptions}, three subjects (02008, 02021, 02023) substructure descriptions in multi-language style, i.e., with annotated language, and by tags like \underline{p}referred name, \underline{d}escription, de\underline{f}inition or \underline{n}ame. Regarding example values, we found 4 forms of indicating the language of a \textit{multi-lang}uage property (Table~\ref{tab_subjects} indicates the formats/separators with \textit{l} as language placeholder) while 4 specifications (marked with "x") denote multi-language examples without any language indication. Further, regarding numerical examples, a subset of the \idtaSpecs{} uses 4 different forms of stating a \textit{unit} after a value (in Table~\ref{tab_subjects} illustrated with \textit{u} as unit placeholder), while the PCF specification/draft (marked by "*") state units after the value. 

If we assume for each category one of the variants in Table~\ref{tab_subjects} as intended, we can estimate more than 570 combinations of unintended variants. Besides those variants indicated in Table~\ref{tab_subjects}, we identified further 12 varying aspects, among them how parent types are specified, how the submodel element class is stated, how example values are separated, whether footnotes are used in table cells, or how value types are stated. This allows for more than 5 billion combinations of writing and formatting variants that our approach is confronted with.

In the following sub-sections, we evaluate our approach with the characterized subjects, i.e., the PDF-based transformation in Section \ref{sect-eval-pdf}, an experiment to tolerantly process the PDF specifications with AI in Section \ref{sect-eval-AI}, the AASX-based transformation in Section \ref{sect-eval-aasx} and the subsequent code generation in Section \ref{sect-eval-gen}. As we apply prototypes to evolving specifications, we discuss limitations in Section \ref{sect-limitations}.

\subsection{PDF-based processing}\label{sect-eval-pdf}

The PDF-based processing can successfully parse all intermediary Excel files obtained from SmallPdf for all subjects and transform them to correct IVML model files. Thereby, the processing also correctly identifies and extracts \textit{enumerations} for 10 specifications, which allow for more specific API generation (cf.~Table \ref{tab_pdf}), including open, extensible enumerations. 

A manual element-by-element comparison between the derived IVML models and the underlying specifications confirmed the correctness of the results, but also revealed issues. The two authors independently evaluated the results for all specifications and agreed on the results in a consensus meeting. 12 \idtaSpecs{} (52 cases) contain \textit{specification issues}, i.e., syntax errors or omitted information such as value types. In some cases we can safely assume defaults, in other cases we can fix the issue by resolving the respective \semanticId{} during postprocessing. These issues do not cover \textit{type issues}, in particular variants of submodel element type names (we found 15 variants including typos) or value types (28 variants including typos, e.g, 9 variants for \texttt{Integer}), which we tolerantly map to a corresponding unified type. Further, two specifications (02010, 02021) use modified type names that do not occur in any other \idtaSpec{}.

Due to the high number of format variants as well as additional whitespaces introduced during the Excel preprocessing, we employed several parsing heuristics. As we feared that heuristics, which could handle the identified issues, could conflict with already implemented heuristics, we manually \textit{modified} the Excel input for 9 specifications, i.e., in 82 cases. The majority of the modifications address accidentally broken \semanticId{} URLs. Further, for 7 specifications (53 cases) our approach does not correctly/completely extract example values (\textit{example issues}). Here, we neither considered new heuristics nor fixed the issues  manually, as we use the results only in tests (\ref{req-test}) and rely there on mechanisms, which tolerantly determine values from the extracted examples or use default values as fallback. In both cases, syntactic improvements of the PDF specifications or alternative formatting like additional table rows/columns or explicit separator tokens could resolve most issues at their root causes.

\begin{table*}[htbp]
\centering
\caption{PDF-based processing of the subjects}
\setlength{\tabcolsep}{0.5em} 
\begin{tabular}{ |l|c|c|c|c|c|c|c|c|c|c|c|c|c|c|c|c|c|c|c| } 
\hline
\textbf{Aspects}&\rotatebox[origin=c]{90}{\textbf{02001}}&\rotatebox[origin=c]{90}{\textbf{02002}}&\rotatebox[origin=c]{90}{\textbf{02003}}&\rotatebox[origin=c]{90}{\textbf{02004}}&\rotatebox[origin=c]{90}{\textbf{02005}}&\rotatebox[origin=c]{90}{\textbf{02006}}&\rotatebox[origin=c]{90}{\textbf{02007}}&\rotatebox[origin=c]{90}{\textbf{02008}}&\rotatebox[origin=c]{90}{\textbf{02010}}&\rotatebox[origin=c]{90}{\textbf{02011}}&\rotatebox[origin=c]{90}{\textbf{02012}}&\rotatebox[origin=c]{90}{\textbf{02013}}&\rotatebox[origin=c]{90}{\textbf{02014}}&\rotatebox[origin=c]{90}{\textbf{02015}}&\rotatebox[origin=c]{90}{\textbf{02016}}&\rotatebox[origin=c]{90}{\textbf{02017}}&\rotatebox[origin=c]{90}{\textbf{02021}}&\rotatebox[origin=c]{90}{\textbf{02023}}&\rotatebox[origin=c]{90}{\textbf{2023-01-24}}\\ 
\hline
\textbf{enumerations}&0&4&0&1&0&0&0&1&2&1&0&1&6&0&0&0&3&6&6\\
\hline
\textbf{specification issues}&3&0&0&4&0&1&0&21&1&0&3&0&0&4&5&12&2&1&2\\
\textbf{type issues}&0&0&0&0&0&0&0&0&3&0&0&0&0&0&0&0&55&0&0\\
\hline
\textbf{modified}&3&0&0&0&0&5&0&0&0&3&26&0&0&0&1&27&7&1&12\\
\textbf{example issues}&0&0&3&2&0&1&2&0&0&0&5&0&0&0&0&5&40&0&0\\
\hline
\end{tabular}
\label{tab_pdf}
\end{table*}

\subsection{AI-based PDF processing} \label{sect-eval-AI}

AI could be a rescue to process input with unintendedly varying syntax more tolerantly. To test the idea, we ran a small experiment with Large Language Models (LLMs) on our subjects. We tried to directly feed the IDTA PDFs into openAI assistant. However, as our account is limited, we were not able to choose the more recent GPT-4 models, and, due to a rate limitation correlating with the input file size, we did not get outputs. Instead we utilised GPT-3.5 Turbo and GPT-4 Turbo through a university provided frontend on the plain texts extracted with Py2PDF. From that, we tried to get the property type definitions as a formatted table or as a JSON structure for further processing, hoping to eliminate the variability in notation and information provided across the \idtaSpecs{}. 
We opted for a one-shot approach by giving the model an example of what we expect to extract from the text. The LLM prompt consisted of a request to turn all tables in the input into a JSON format like the given example with constraints that the LLM shall not utilise information from the example or add any additional explanations to the output JSON. We ran each prompt three times to test for variations in the answers. While the output was mostly correct JSON, we found that in the results data was occasionally missed or hallucinated while asking the same prompt multiple times would produce different outputs. Our conclusion is that currently the employed general models cannot be used in our automated process without massive human intervention.

\subsection{AASX-based processing} \label{sect-eval-aasx}

At a glance, a machine-readable format might be the better choice as input for a model transformation. However, also our AASX transformer has to cope with variants and we identified several issues in the extracted information, justifying our PDF-based approach. One major variant, as indicated in Table \ref{tab_aasx}, is the \textit{AASX version}, which also seems to require different XML element names. An example for a more lower-level variant is how the AASX files denote \textit{cardinalities}: 50\% use the qualifier "\underline{c}ardinality", 38\% the qualifier "\underline{m}ultiplicity" and the remaining 11\% completely omit the information although present in the \idtaSpec{}. Similarly, element names or sub-structures differ, e.g., how to locate property descriptions/example values or which structural schema to apply for concept descriptions. Further, Table \ref{tab_aasx} indicates whether \textit{BaSyx} (in for relevant versions 1.\underline{0}, 1.\underline{3}) was able to read an AASX file.

44\% of the AASX files state the \textit{specification version} they are targeting. However, for 02007 the version number seems to be wrong and half of the AASX files do not state a version at all. Moreover, not only names and structures, but also the contents differs significantly with respect to the targeted \idtaSpec{}. As in Section \ref{sect-eval-pdf}, we independently compared the output of our model transformation with the respective \idtaSpec{} and validated issues in the respective AASX file. Only 22\% of the AASX files include \textit{notes for \ref{req-addInfo}} stating additional information, i.e., we are missing more than 35 relevant notes from the PDFs. In 38\% we found that the \textit{\idShort{}} of properties, and in 72\% the \textit{type} of a property is differing with respect to the \idtaSpec{}. This is particularly evident for the sizing of power trains (02021) and the asset interface (02017) AASX and correlates with the specification issues in Section \ref{sect-eval-pdf}. Except for reliability (02013) and safety (02014), there are almost no missing \textit{cardinalities} and only some cardinalities differ with respect to the PDF. Also most \textit{\semanticIds{}} are correct, but 5 are missing in 02012 and some are differing in the AASX for 02017 and 02021. Further, for 02012 and 02017 almost all \textit{descriptions} are missing, while in 02017 several descriptions are either missing or significantly differing. Moreover, in the AASX files for 02008, 02014, 02017, 02021 and 02023 we found missing \textit{examples}, while many are differing in the AASX for 02006. 

As far as we observed, several issues arise from the purpose an AASX file was created for. If a template for a specification is represented, the contents tends to be closer to the specification and the number of issues tends to be lower compared to instantiated examples (instead of templates) or even a mix of templates and examples. We expressed this in Table \ref{tab_aasx} by the numbers of \textit{instantiated}, \text{missing} (but expected) and \textit{additional} types/fields, which correlate in particular for 02021, 02017 and 02008 with a higher number of issues as discussed above. 

To further quantify the differences, we also estimated the \textit{overlap} of the resulting IVML models for PDF and AASX-based transformation. Here, we looked for equality of \idShorts{}, types and \semanticIds{}, while for description and examples we counted equality already if they are present in both output models. We found the highest overlap of 88\% for specification 02006 and the smallest overlap for 02013. However, depending on the size/complexity of the specifications, already a few issues may lead to larger differences in the overlap.

\begin{table*}[htbp]
\centering
\caption{AASX-based processing of the subjects (no AASX for PCF draft 2023-01-24).}
\setlength{\tabcolsep}{0.5em}
\begin{tabular}{ |l|c|c|c|c|c|c|c|c|c|c|c|c|c|c|c|c|c|c| } 
\hline
\textbf{Aspects}&\rotatebox[origin=c]{90}{\textbf{02001}}&\rotatebox[origin=c]{90}{\textbf{02002}}&\rotatebox[origin=c]{90}{\textbf{02003}}&\rotatebox[origin=c]{90}{\textbf{02004}}&\rotatebox[origin=c]{90}{\textbf{02005}}&\rotatebox[origin=c]{90}{\textbf{02006}}&\rotatebox[origin=c]{90}{\textbf{02007}}&\rotatebox[origin=c]{90}{\textbf{02008}}&\rotatebox[origin=c]{90}{\textbf{02010}}&\rotatebox[origin=c]{90}{\textbf{02011}}&\rotatebox[origin=c]{90}{\textbf{02012}}&\rotatebox[origin=c]{90}{\textbf{02013}}&\rotatebox[origin=c]{90}{\textbf{02014}}&\rotatebox[origin=c]{90}{\textbf{02015}}&\rotatebox[origin=c]{90}{\textbf{02016}}&\rotatebox[origin=c]{90}{\textbf{02017}}&\rotatebox[origin=c]{90}{\textbf{02021}}&\rotatebox[origin=c]{90}{\textbf{02023}}\\ 
\hline
\textbf{AASX version}&2&2&2&2&2&2&2&2&2&2&2&2&2&2&2&2&3&3\\
\textbf{cardinalities}&c&m&c&c&m&m&m&c&c&m&c&-&-&m&m&c&c&c-\\ 
\textbf{BaSyx}&0/3&-&0&0/3&0/3&-&-&0/3&-&0/3&-&0/3&0/3&0/3&0/3&0/3&-&-\\
\hline
\textbf{spec version}&1.0&-&1.2&1.2&-&-&0.15&1.1&-&-&-&1.0&1.0&-&-&-&1.0&0.9\\
\textbf{notes for \ref{req-addInfo}}&-&x&-&x&-&x&x&-&-&-&-&-&-&-&-&-&-&-\\
\textbf{idShort diff}&0&0&0&0&0&3&2&0&1&0&5&0&1&0&0&1&2&0\\
\textbf{type diff}&0&0&3&2&0&1&2&2&2&0&2&1&0&4&1&11&53&4\\
\textbf{cardinality no/diff}&0/1&0/0&0/0&0/0&0/0&1/0&0/3&0/0&0/1&0/0&0/2&14/0&31/0&0/0&0/0&2/5&0/1&0/0\\
\textbf{\semanticId{} no/diff}&0/0&0/1&0/2&0/0&0/0&0/2&0/2&0/1&0/0&0/0&5/22&0/2&0/1&0/2&0/0&0/7&0/5&0/2\\
\textbf{description no/diff}&2/6&0/6&0/1&0/5&3/0&3/17&1/18&0/5&4/8&0/14&55/0&1/6&0/37&1/0&0/5&80/4&12/26&0/2\\ 
\textbf{example no/diff}&0/4&0/3&4/2&2/2&0/0&0/28&6/9&32/0&12/0&0/0&5/0&10/0&26/0&11/0&7/0&52/2&167/2&16/0\\
\hline
\textbf{instantiated}&2&0&0&0&0&0&0&1&0&0&5&0&0&0&0&12&0&1\\
\textbf{missing/additional}&1/12&0/0&0/2&0/2&0/0&0/0&0/2&1/5&10/0&0/0&1/0&0/1&0/0&5/3&0/4&22/1&4/0&0/0\\
\hline
\textbf{overlap}&52\%&81\%&81\%&78\%&76\%&88\%&55\%&66\%&60\%&74\%&60\%&39\%&41\%&67\%&69\%&41\%&66\%&66\%\\
\hline
\end{tabular}
\label{tab_aasx}
\end{table*}

\subsection{API Code Generation}\label{sect-eval-gen}

For evaluating the code generation, we focus now on the IVML models from the PDF-based transformation, as they better represent the specifications. The approach successfully creates per submodel specification two documented Java API classes (AAS creation and accessor) as well an associated unit test. For smaller submodels like the control component type (02015) or the control component instance (02016) about 300 Commented Lines Of Code (CLOC) are generated. For the larger ones, e.g., the digital nameplate (02006) we counted 5000 CLOC and for the sizing of power trains submodel (02021) even more than 11000 CLOC. The size of the test cases for the smaller submodels (02015, 02016) is about 150 lines, while for 02006 more than 900 and for 02021 more than 1900 CLOC are generated. A manual inspection showed that all relevant structures and properties are present. All generated Java are validated by running their generated build process, i.e., compilation and unit test can be executed successfully. The average line/instruction coverage of the tests is 87\%. It is important to emphasize that in particular the hierarchical structure submodel (02011) and the asset interfaces description submodel (02017) contain direct and indirect structural recursions, which are correctly handled by the generated tests. 

\subsection{Limitations}\label{sect-limitations}

We aimed for an agile realization of a proof-of-concept prototype, which is neither meant to be complete nor ready for production. In more details, not all primitive AAS types are considered, AAS operations are left out and more recent AAS types like lists are currently mapped to collections. Further, analyzing recent IDTA specifications limits generalizability as we focus on a certain point in time. Moreover, issues identified in this work will hopefully disappear in the next versions of the specifications, i.e., may impact the long-term validity of heuristics, alternatives and results in our approach. In addition, if further AAS elements and formats occur, as, e.g., lists, our approach needs to be adjusted and extended. 

We are aware of these limitations and intentionally pursued an early proof-of-concept realization and an evaluation based on the actual state to also derive lessons learned and suggestions. 
In the future, advanced model mechanisms could (at a glance) supersede our approach, e.g., if a submodel template could be directly instantiated within an AAS framework. We believe that our code generation approach is then  still useful as dedicated, typed creation operations would still be a desirable API compared to generic framework calls. Although the creation API might then become simpler, the typed access API as well as the tests may then still be similar.

\section{Discussion and Lessons Learned}\label{sect-lessons}

In this section, we summarize the experiences that we made when realizing and evaluating our approach, first focusing on the PDF specifications then on the AASX files.

In an evolving ecosystem, \textbf{versioning} is mandatory, for specifications, AASX files, imported specifications and the required meta-model. While all PDF files state a version and most \semanticIds{} are versioned, only 44\% of the AASX files explicitly indicate the version of the targeted specification.
Further, except for literature references in \idtaSpecs{} pointing to the meta-model, only the technical data specification (02003) explicitly states the required AAS meta-model version. 

The current \idtaSpecs{} contain various \textbf{semantic issues}, ranging from missing value types over wrong submodel element types to discrepancies between PDF and AASX. Creating a semantically consistent specification using a text processor without further tools can be compared to programming without an IDE or a style checker. Here, our PDF-based model translation shows that tools for checking semantics and presentation guidelines can be realized. Identified technical issues like additional whitespaces could be resolved using the actual source as input, e.g., a word file, rather than a print format like PDF. This would then also prevent the formatting issues introduced by our Excel detour. To further increase consistency, one could directly derive AASX templates from the specification document, thus, avoiding the identified AASX issues and eliminating the effort for manual validation. In our approach, this can be achieved by adding an AASX output formatter. If the creation with external AAS tools is desired, parsing the AASX  and comparing the resulting model with the one for the specification could ease the validation.

The PDF specifications contain \textbf{notes that may imply constraints}, e.g., whether a name of a submodel is fixed or which (enumerated) values or \semanticIds{} are valid for a certain property. However, as not all \idtaSpecs{} include such notes, it remains unclear to us whether the given notes shall be considered mandatory. Here uniform semantics and presentation, e.g., as explicit constraints, would be desirable. As the specifications and the AAS meta-model employ OMG UML for illustrating the structure of submodels, we could imagine that the OMG Object Constraint Language (OCL) could be used to unambiguously specify constraints.

In the PDF tables, several \textbf{aspects are mixed in the same cell}, e.g., \idShort{} with examples, \semanticId{} with description or  sometimes example(s) with example explanations. Varying presentation forms are an obstacle for human understanding and programmatic parsing. Clear and uniform formats and separators, e.g., as further table rows or in terms of syntax would be desirable. As for semantic checking, a more strict version of our PDF parser could be used as style checker.

While older specifications tend to be repetitive, more recent specifications utilize \textbf{mechanisms for reuse}. While our PDF-based model extraction implements a feasible interpretation, a uniform set of mechanisms as well as clear semantics are needed to prevent issues and to ensure interoperability. Further, as we experienced, the aspect-like notation in more recent \idtaSpecs{} is difficult to read and understand for humans.

Most of our observations so far focus on the PDFs. In general, it is important that relevant information including  constraints and additional information (\ref{req-addInfo}) is \textbf{equally represented in AASX templates} and, thus, machine-readable.  Moreover, we consider \textbf{mixing submodel templates with examples} as problematic, in particular if not the full template mandated by a specification is in the AASX file. However, we value the effort of creating examples for specifications, for illustration as well as for validation purposes.

\section{Related Work}\label{sect-related}

In this section, we briefly review related work on the main topics of this paper, namely data integration/format conversion involving AAS, model transformations for AAS and AI as enabler for tolerant format parsing. 

Several approaches aim at \textbf{integrating data} into or translating data for AAS. For example, Ye\etal{} use in~\cite{YeSongWon+22} Excel files as intermediary format for bi-directional data exchange between AAS and enterprise applications. As an enabler, the AasTransformation library\footnote{\url{https://admin-shell-io.github.io/aas-transformation-library/}} allows for the creation of AAS instances from XML, in particular OPC UA or AutomationML. While these approaches focus on the transfer of data in different formats into/with AAS, we primarily take over specification examples into our models to generate tests.

On the \textbf{meta-model level}, Platenius-Mohr\etal{} consider in~\cite{PMMG+20} the use case of a customizable bi-directional mapping between AAS. Schmidt\etal{} propose a model-transformation between the Digital Twin Definition Language and AAS in~\cite{SchmidtVolzStojanovic+23}. Lüder\etal{} generate in~\cite{LuederBehnertRinker+20} AAS from AutomationML, while  Cavalieri and Salafia in~\cite{CS22} as well as Weiss and Reichelt in~\cite{WeissReichelt23} discuss model-based mapping/transformation between \OPCUA{} and AAS. As a potential enabler, Miny\etal{} propose in~\cite{MTE+20} a custom-build model-transformation language based on OMG UML and OCL. While these approaches focus on transformations between models providing similar data, we use an augmenting model-transformation between the AAS meta-model and IVML to automatically generate AAS APIs/tests. 

In case of syntactically differing input files, \textbf{AI-based information extraction} may be a solution. Large language models can support the data extraction from files. In~\cite{ni2023chatreport}, ChatGPT and GPT-4.0 are used to analyze sustainability reports with around 70 pages. The authors found that ChatGPT provides better results as its answers are more focused on the input report. They utilised prompt engineering to improve the evaluation results. For our use case, which mostly consists of structured tables, we need stably structured outputs from the model. Gao\etal{} \cite{gao2023exploring} explored ChatGPT for event extraction, i.e., requiring structured outputs. By using a positive example in the input prompt they managed to consistently get a structured output fit for processing. However, the general result quality heavily depends on the prompts. So far, not much work aimed for tolerantly parsing structured tables.

\section{Conclusion and Future Work}\label{sect-conclusion}

Specifications are key to industrial systems and Industry 4.0. Asset Administration Shell is a new standard in this field and the IDTA submodel specifications aim at structuring and standardizing relevant information. In this context, we asked whether the increasing number of IDTA specifications can be automatically turned into supporting API code. The answer is yes, but problems arise from the input sources at the beginning of the process. Both, PDF and AASX specification representations can be turned automatically into an intermediary, augmented model, whereby ironically the PDFs are currently the better input for model-based approaches. Although we tried, we were not able to compensate current (unintended) syntactic variations through AI so that in some cases manual correction of the input was required. From the intermediary model, our code generation creates more than 50 KLOC API code and 8 KLOC test code for 18 specifications and one draft. Our lessons learned target issues and unintended variations in the specifications, arguing that automated style-checking and model transformations can be used to improve both, specifications and (generated) AASX files.

In the future, we plan to continue evaluating new IDTA specifications with our approach, hoping that after maintenance releases of the current IDTA submodel specifications many of the reported issues will disappear and our technical compensation mechanisms will become superfluous. Further, we plan to integrate the generated models and API code with our \oktoflow{} platform and to validate them in demonstrators.

\section*{Acknowledgment}

We are grateful to the SmallPdf support team for discussing issues on initially corrupted PDFs.

\bibliographystyle{IEEEtran}
\bibliography{ETFA24}
\vspace{12pt}

\end{document}